\documentclass[runningheads]{llncs}
\usepackage[T1]{fontenc}
\usepackage[utf8]{inputenc}
\usepackage{amsmath}
\usepackage{booktabs}
\usepackage{CJKutf8}
\usepackage[dvipsnames]{xcolor}

\usepackage{graphicx}
\usepackage{hyperref}
\usepackage{color}

\usepackage{subcaption}

\definecolor{c1}{HTML}{E85642}

\begin{document}
\title{How role-play shapes relevance judgment in zero-shot LLM rankers}
%
%
\author{
  Yumeng Wang\inst{1}\orcidID{0000-0002-2105-8477} \and
  Jirui Qi\inst{2}\orcidID{0009-0009-0608-8203} \and
  Catherine Chen\inst{3}\orcidID{0009-0009-8734-436X} \and
  Panagiotis Eustratiadis\inst{4}\orcidID{0000-0002-9407-1293} \and
  Suzan Verberne\inst{1}\orcidID{0000-0002-9609-9505}
}

\authorrunning{Y. Wang et al.}

\institute{
  Leiden Institute of Advanced Computer Science, Leiden University\\
  \email{\{y.wang,s.verberne\}@liacs.leidenuniv.nl}
  \and
  Center for Language and Cognition, University of Groningen\\
  \email{j.qi@rug.nl}
  \and
  Brown University\\
  \email{catherine\_s\_chen@brown.edu}
  \and
  University of Amsterdam\\
  \email{p.efstratiadis@uva.nl}
}

\maketitle              
\begin{abstract}
Large Language Models (LLMs) have emerged as promising zero-shot rankers, but their performance is highly sensitive to prompt formulation. In particular, role-play prompts, where the model is assigned a functional role or identity, often give more robust and accurate relevance rankings. However, the mechanisms and diversity of role-play effects remain underexplored, limiting both effective use and interpretability.
In this work, we systematically examine how role-play variations influence zero-shot LLM rankers.
We employ causal intervention techniques from mechanistic interpretability to trace how role-play information shapes relevance judgments in LLMs.
Our analysis reveals that (1) careful formulation of role descriptions have a large effect on the ranking quality of the LLM; (2) role-play signals are predominantly encoded in early layers and communicate with task instructions in middle layers, while receiving limited interaction with query or document representations. Specifically, we identify a group of attention heads that encode information critical for role-conditioned relevance.
These findings not only shed light on the inner workings of role-play in LLM ranking but also offer guidance for designing more effective prompts in IR and beyond, pointing toward broader opportunities for leveraging role-play in zero-shot applications.
\end{abstract}

\section{Introduction}

Large language models (LLMs) have demonstrated a remarkable ability to follow natural language instructions, offering new paradigms for information retrieval (IR)~\cite{abbasiantaeb2024can}: an LLM receives a query, a set of candidate documents, and an instruction describing the ranking criterion; it then produces either a relevance judgment for a single document (pointwise~\cite{liang2022holistic,nogueira2020document,zhuang2023beyond,zhuang2021deep}) or a ranking over multiple documents (pairwise~\cite{qin2023large}, setwise~\cite{zhuang2024setwise}, listwise~\cite{ma2023zero,sun2023chatgpt,tang2023found})\cite{zhuang2024setwise}. 

Previous work shows that LLM rankers are highly sensitive to prompt wording~\cite{sun2025investigation}. Among diverse prompting strategies, \emph{role-play} prompts, which assign the model a functional identity in addition to task instructions (e.g., ``You are RankGPT, an intelligent assistant that can rank passages based on their relevance to the query''~\cite{sun2023chatgpt}), often achieve strong and relatively robust gains. Conceptually, we distinguish between task instruction (what the model should do) and role-play (the capacity or identity under which it should do it).

Prior work typically conducts a simple with-or-without comparison of the prompt with role-play phrasing versus a non-role-play baseline~\cite{sun2025investigation}. This leaves open an important question, 
especially given the increasing prevalence of role-play as a common prompting strategy~\cite{kong-etal-2024-better}:
\begin{itemize}
    \item[] \textbf{RQ0:} How sensitive are LLM rankers to variations in role-play wordings?
\end{itemize}
Precise control over the linguistic surface form of role-play phrasing is required to study this core question. Therefore, we introduce a modular role-play template whose interchangeable lexical slots allow us to generate positive and negative role variants while holding the remainder of the prompt fixed. 
The experimental results show that seemingly minor wording changes to the role description yield large shifts in zero-shot ranking performance, highlighting the need to understand how role information is represented and exploited inside the model.

To bridge this gap, we approach the problem from the perspective of mechanistic interpretability, which seeks to reverse-engineer the internal weights and representations of neural models into human-understandable algorithms.
Specifically, we use activation patching~\cite{meng2022locating,vig2020investigating}, a controlled intervention technique that replaces intermediate model activations with those from alternative prompts, allowing us to trace where and how information is represented and transmitted within the network. This allows us to address the following research questions:
\begin{itemize}
  \item[] \textbf{RQ1:} How is role-play information represented and propagated across layers in zero-shot LLM rankers?
  \item[] \textbf{RQ2:} Are there attention heads whose causal effects on generation are specifically tied to role-play signals?
  \item[] \textbf{RQ3:} Which linguistic components of a role-play prompt contribute most to guiding zero-shot LLM rankers?
  \item[] \textbf{RQ4:} How stable are the role-play effects revealed by activation patching across models and datasets?
\end{itemize}
Using the designed role-play template, we construct counter-prompt pairs and apply activation patching to localize the model components responsible for encoding role-play signals, as well as to trace how these signals interact with instructions and relevance judgments throughout the network.

Taken together, these findings clarify why role-play prompting can be so effective for zero-shot ranking and offer actionable guidance for designing robust prompts and for developing architectures or training recipes that explicitly encode beneficial role signals. Our contributions are fourfold: (1) We systematically evaluate role-play prompts for pointwise and pairwise LLM rankers, showing substantial performance differences between positive roles and negative roles; (2) through activation patching, we demonstrate how role-play information propagates through the model, interacts modestly with instruction tokens;
(3) we find that ablating attention heads identified via patching can substantially alter model behavior, indicating that these heads encode information critical for role-conditioned relevance;
(4) we demonstrate that role-play information propagation in zero-shot LLM rankers is generally consistent across models and datasets. Our code is available at \url{https://github.com/menauwy/roleplay_zeroshot}.
\section{Related work}

\subsubsection{Zero-shot LLM ranking}
Several zero-shot ranking methods based on LLMs have recently been proposed, falling into four categories: pointwise~\cite{zhuang2023beyond,zhuang2021deep,zhuang2021tilde}, pairwise~\cite{qin2023large}, listwise~\cite{ma2023zero,sun2023chatgpt,tang2023found}, and setwise~\cite{zhuang2024setwise}. Pointwise methods~\cite{liang2022holistic,nogueira2020document} prompt the LLM with a query–document pair and ask whether the document is relevant, often generating a binary answer (``Yes''/``No''); the likelihood of ``Yes'' is used for ranking. Pairwise methods~\cite{qin2023large} compare two documents for a query, ranking based on relative preference. Listwise methods~\cite{ma2023zero,sun2023chatgpt,tang2023found} prompt the LLM to label a list of documents in order, while setwise methods~\cite{zhuang2024setwise} iteratively select the most relevant documents from a set until the top-k are returned.
We particularly focus on the pointwise setting for the relevance judgment task and the pairwise setting for the ranking task.


\subsubsection{Mechanistic interpretability in IR}
Mechanistic interpretability~\cite{cunningham2023sparse,elhage2021mathematical,wang2023interpretability} seeks to understand how specific components of neural networks drive model behavior. A key method is activation patching~\cite{meng2022locating} (also called causal tracing~\cite{meng2022locating} or causal mediation analysis~\cite{vig2020investigating}), which manipulates intermediate activations to trace information flow and localize behavior to components such as attention heads. 
Existing interpretability IR methods generally lack of such fine-grained analysis~\cite{anand2022explainable}, with a few studies exploring it.
In neural IR, Chen et al.~\cite{chen2024axiomatic} use this technique to isolate components satisfying term-frequency axioms. In zero-shot LLM ranking, Liu et al.~\cite{liu2025large} apply it to study how various LLM modules influence pointwise and pairwise relevance predictions. Building on this line of work, we investigate how role-play shapes relevance judgments in LLM rankers.


\subsubsection{Prompt variations}

Prompt variations have been studied in the context of zero-shot LLM ranking~\cite{sun2025investigation}. Sun et al.~\cite{sun2025investigation} show that role-play makes a positive difference to the retrieved results, although they examine only one variant of role-play (``RankGPT'') against a non-role-play baseline.
Our work further draws inspiration from Kong et al.~\cite{kong2023better} who study role-play prompting in the context of reasoning tasks. Their results show 
that LLMs that were told ``You are an expert at math'', perform better at mathematical reasoning than if not told. This suggests that models have internal knowledge that allows them to answer questions correctly, but do not unless explicitly prompted. We explore the implications of this phenomenon in information-seeking tasks.
 \section{Methods}

\subsection{Zero-shot LLM ranking with role-play}



\begin{figure}[t]
    \centering
    \includegraphics[width=\linewidth]{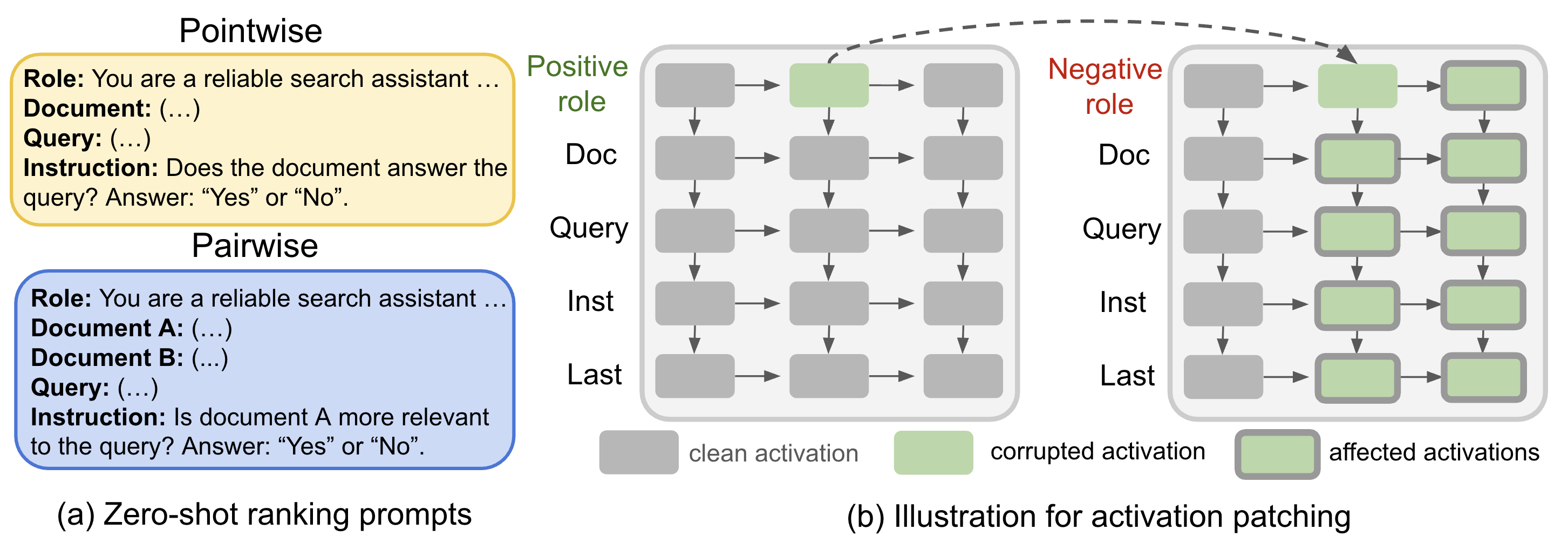}
    \caption{Role-play activation patching for zero-shot rankings.}
    \label{fig:pipeline}
\end{figure}

Zero-shot LLM-based ranking leverages the instruction-following capabilities of LLMs to assess query-document relevance via prompting.
In this work, we consider both pointwise and pairwise zero-shot settings. 
In the pointwise setting, the model assesses one query-document pair at a time, which aligns naturally with relevance judgment tasks~\cite{zhuang2023beyond}.
In the pairwise setting, each query is evaluated against two documents, a design that naturally encodes a relative relevance comparison and is therefore well-suited for ranking tasks~\cite{qin2023large}.
We exclude listwise and setwise approaches primarily because their multi-document context introduces intricate cross-document attention dependencies that would complicate the causal tracing of internal mechanisms. Furthermore, such settings are less amenable to a constrained and unified output space.
To maintain consistency between pointwise and pairwise prompts, we unify the output format to ``Yes'' or ``No'' and structure each prompt according to the template illustrated in Figure~\ref{fig:pipeline}a. For pairwise prompts, we control for positional bias by ensuring that Document A appears equally often as relevant and irrelevant across samples.
Given the high computational cost of activation patching, we restrict each sample query to a single document (pointwise) or a single document pair (pairwise) and report results averaged over all samples.

\subsubsection{Role-play prompt design}\label{sec:prompt_design}
To disentangle role-play from task instruction, we deliberately place the role-play segment at the beginning of a prompt and the instruction at the end. We design a flexible template that always tokenizes to the same number of tokens, which enables reliable activation patching: \textit{You are a/an \{adjective\} search assistant that \{modal verb\} rank passages \{adverb\}, based on their relevance to a query.}
For the adjective and adverb slots, we select 10 distinct positive tokens and 10 distinct negative tokens each, while 3 tokens are selected for the modal slot. The selected tokens are shown in Table~\ref{tab:lexical_slots}. This results in $10 \text{(adjective)} \times 3 \text{(modal)} \times 10 \text{(adverb)} = 300$ unique combinations of possible role-play prompts, either positive or negative, which we use for RQ0 to evaluate zero-shot ranking performance during inference.
For the activation patching experiments, we select 10 positive-negative prompt pairs with completely distinct lexical slots and no repetitions except modal words.
\begin{table}[ht]
\centering
\footnotesize
\caption{Lexical slots used in role-play prompts, organized by type and polarity.}
\label{tab:lexical_slots}
\begin{tabular}{l p{5cm} p{5cm}}
\toprule
\textbf{Slot} & \textbf{Positive} & \textbf{Negative} \\
\midrule
Adjective & talented, expert, superb, capable, reliable, gifted, brilliant, focused, clear, knowledgeable& faulty, confused, clumsy, sluggish, incorrect, awful, hopeless, flawed, problematic, unreliable \\
\midrule
Adverb & carefully, correctly, swiftly, wisely, perfectly, accurately, nicely, fairly, logically, clearly & wrongly, poorly, mistakenly, slowly, falsely, terribly, badly, incorrectly, sadly, horribly \\
\midrule
Modal& can, will, shall & can, will, shall \\
\bottomrule
\end{tabular}
\end{table}


\subsection{Activation patching}\label{sec:actpat}

Activation patching~\cite{meng2022locating,vig2020investigating}
provides a principled way to analyze whether specific model components are causally involved in producing a target behavior. The central idea is to intervene on latent activations: given a clean prompt that elicits the desired output and a corrupted prompt that disrupts it, we replace the corrupted activations at a chosen component with cached values from the clean run, and then evaluate whether the model recovers its original behavior. In this sense, the method serves as a causal probe of the model's internal computation.

In our setting, the behavioral contrast comes from role-play manipulation. A clean prompt instantiates a positive role (e.g., \textit{You are a capable search assistant...}), while the corrupted counterpart instantiates a negative role (e.g., \textit{You are an unreliable search assistant...}), as illustrated in Figure~\ref{fig:pipeline}b. For both pointwise and pairwise ranking prompts, the expected label under the clean run for a relevant document is ``Yes'', whereas the corrupted run for a relevant document is expected to produce ``No''. By patching activations from the positive-role run into the negative-role run, we can test whether certain components are responsible for restoring the model's ability to recognize relevance under the intended role.
Formally, activation patching proceeds in three stages~\cite{heimersheim2024use,zhang2023towards}:
\begin{itemize}
    \item[] \textbf{1. Clean run:} Run the model on a clean (positive-role) prompt and cache activations from the selected component(s) at specific token positions.
    \item[] \textbf{2. Corrupted run:} Replace the role with a negative version and run the model, recording the predicted output.
    \item[] \textbf{3. Patched run:} Re-run the corrupted prompt, but replace its activations at the chosen locations with those cached from the clean run. The causal contribution of the component is measured by how much the model’s output shifts toward the clean prediction.
\end{itemize}
We conduct patching at multiple levels of model components: the residual stream, attention blocks, and MLP blocks, with a finer-grained analysis at the level of individual attention heads within each layer.
This design allows us to localize which components and positions most strongly mediate the role-play effect in zero-shot ranking. 

\subsubsection{Metrics}

Following related work in activation patching~\cite{polyakov2025towards,zhang2023towards}, we use logit difference (LD) to quantify the model's preference for ``Yes'' over  ``No'' when ``Yes'' is the correct answer.
Denoting the logits as $\ell$, and the vocabulary index of tokens ``Yes'' and ``No'' as $i_{\text{Yes}}$ and $i_{\text{No}}$ respectively, we write 
$\mathrm{LD} = \ell_{i_{\text{Yes}}} - \ell_{i_{\text{No}}}$, 
Using correct, corrupted, and patched runs, we can obtain the normalized LD following~\cite{wang2023interpretability} as:
\begin{equation}
    \text{normalized LD} = \frac{\text{LD}_{\text{patched}}-\text{LD}_{\text{corrupted}}}{\text{LD}_{\text{clean}}-\text{LD}_{\text{corrupted}}}.
\end{equation}
Its range is $[0,1]$, with 0 indicating no recovery from patching and 1 indicating full performance restoration.
While we use ``Yes'' for illustration, the formulation applies equally when the correct answer is ``No'', such as for positive roles paired with irrelevant documents and negative roles paired with relevant documents.

\section{Experimental setup}
\subsubsection{Model}
We use instruction-tuned LLMs, as their strong ability to follow instructions also makes them well-suited to reliably play assigned roles.
For the main experiments, we use LLaMa-3.1-8B-Instruct~\cite{grattafiori2024llama}, chosen for its wide adoption in related work to facilitate fair comparison. 
We adopt a zero-shot setting, as our goal is to examine the raw effect of role-play on model behavior without introducing bias from task-specific supervised finetuning.
We additionally include experiments with other widely used LLMs to assess generalizability.

\subsubsection{Dataset}\label{dataset}
We first assess the sensitivity of zero-shot LLM rankers to role-play wordings using the TREC-DL datasets~\cite{craswell2025overview}. For activation patching, we primarily utilize MS MARCO~\cite{bajaj2016ms} for relevance judgment tasks, while Natural Questions (NQ)~\cite{kwiatkowski-etal-2019-natural} is used to evaluate generalization. Additionally, ranking-specific patching tasks are conducted on TREC-DL.
Following the patching setup in Liu et al.~\cite{liu2025large}, we sample 100 queries from the training set, each paired with one relevant document from the human-provided labels and one irrelevant document, both randomly drawn from the top 100 documents retrieved by BM25~\cite{robertson2009probabilistic}.
For pointwise ranking, we use 50 samples with relevant documents and 50 with irrelevant ones. 
For pairwise ranking, we use 50 samples with \textit{Document A} being relevant and 50 samples with \textit{Document A} being irrelevant. Each \textit{Document A} is paired with a counter-document (\textit{Document B}) of the opposite relevance.

For activation patching, we construct 10 positive-negative prompt pairs using non-repeating lexical slots from Table~\ref{tab:lexical_slots}.
Each prompt pair is applied to all 100 ranking samples, and results are averaged across the 10 pairs for stability. This design balances computational efficiency with robustness, with the observation that even a single pair yields qualitatively similar insights.

\section{Results}

\subsection{LLM rankers sensitivity to role-play wordings (RQ0)}
\begin{figure}[t]
    \centering
    \includegraphics[width=0.8\linewidth]{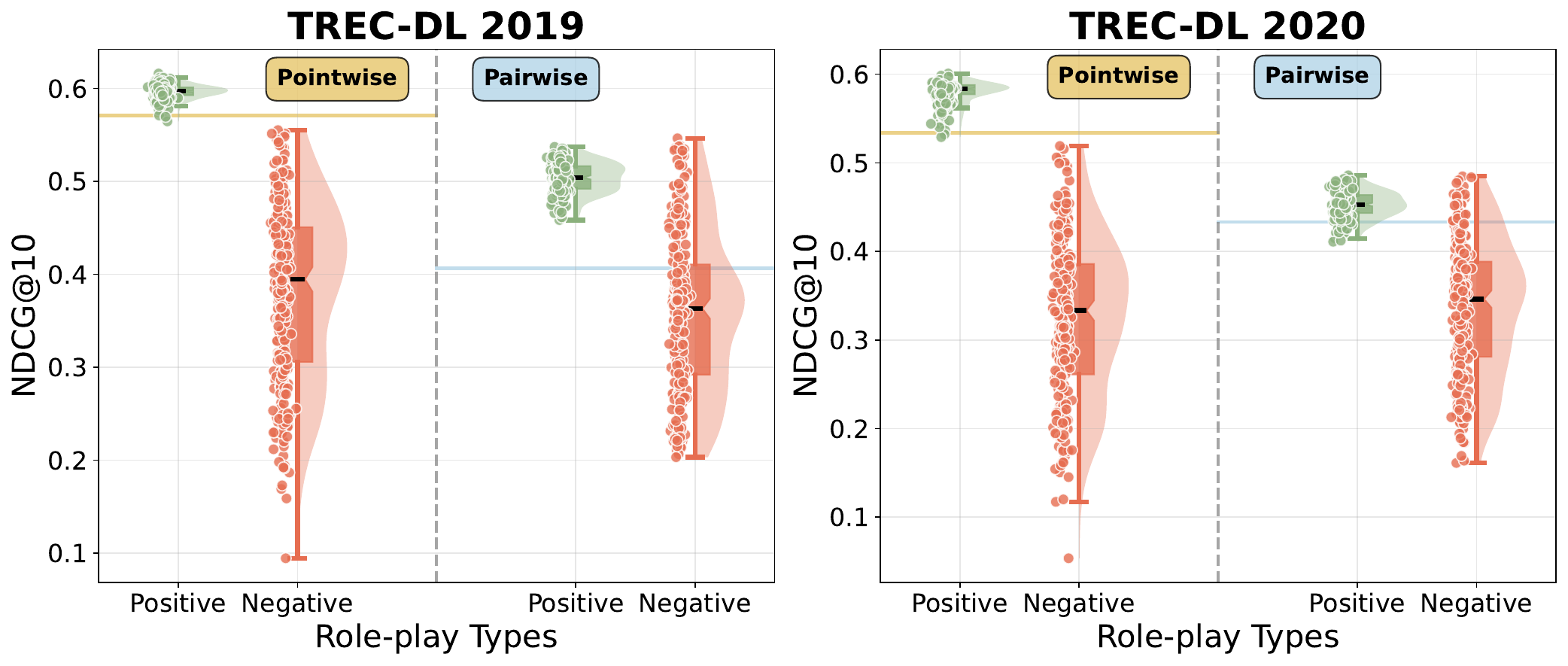}
    \caption{Zero-shot LLM ranker performance (nDCG@10) for various role-play wordings across methods, prompt types, and datasets on LLaMA-3.1-8B-Instruct. Baselines without role-play appear as horizontal lines.}
    \label{fig:rq0_role_sensitivity}
\end{figure}

Building on the findings of~\cite{sun2025investigation}, we extend their experiments to assess the sensitivity of zero-shot LLM rankers to variations in role-play wordings, using the same settings on TREC-DL datasets~\cite{craswell2025overview}. Role-play prompts are constructed following the template introduced in Section~\ref{sec:prompt_design}.
Figure~\ref{fig:rq0_role_sensitivity} summarizes two key findings:

\textbf{Sensitivity to wording} Even small variations in three words can substantially influence ranking performance across role-play types. The effect is especially pronounced for negative role-play prompts: on TREC-DL 2019~\cite{craswell2025overview}, performance ranges from 0.094 to 0.555 depending on the specific wording. In contrast, positive role-play prompts yield more consistent results, which we suspect may be due to the model’s natural tendency to follow task instructions positively even without an explicit functional role.

\textbf{Interaction with ranking methods} 
We assume that positive role-play prompts should improve performance relative to the baseline, while negative prompts should degrade it, such that the performances of positive and negative prompts do not overlap. 
Under pointwise ranking, this expectation holds.
In pairwise ranking, most positive prompts still outperform the baseline, but some perform worse.
Notably, negative prompts often deviate from expectations, occasionally outperforming the non-role-play baseline and in some cases showing performance comparable to or surpassing that of positive prompts.
These observations motivate a deeper investigation into how role information is represented and utilized within the model.

\subsection{Information flow of role-play (RQ1)}

\begin{figure}[t]
  \centering
  \begin{subfigure}{0.49\textwidth}
    \centering
    \includegraphics[width=0.8\textwidth]{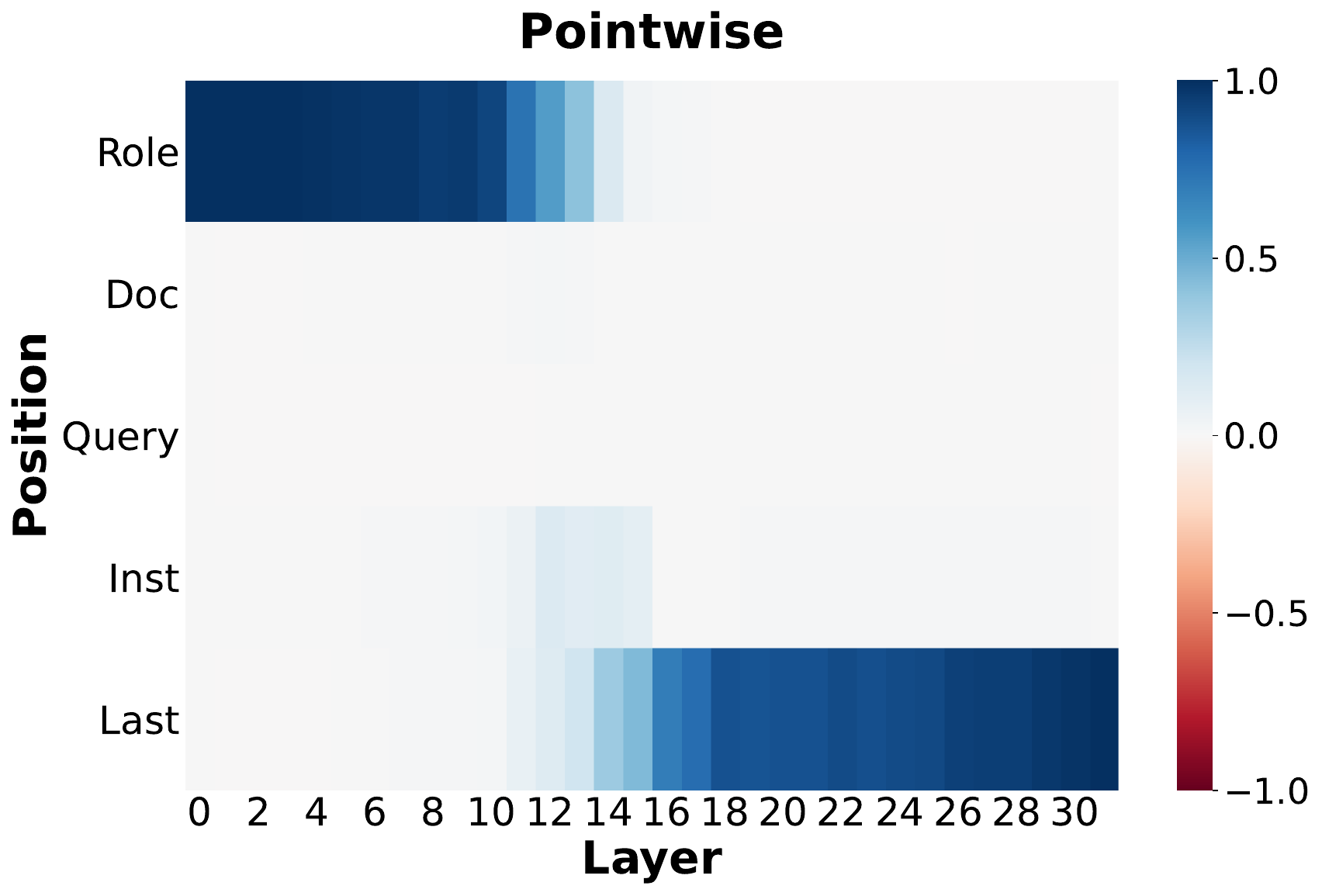}
  \end{subfigure}
  \hfill
  \begin{subfigure}{0.49\textwidth}
    \centering
    \includegraphics[width=0.8\textwidth]{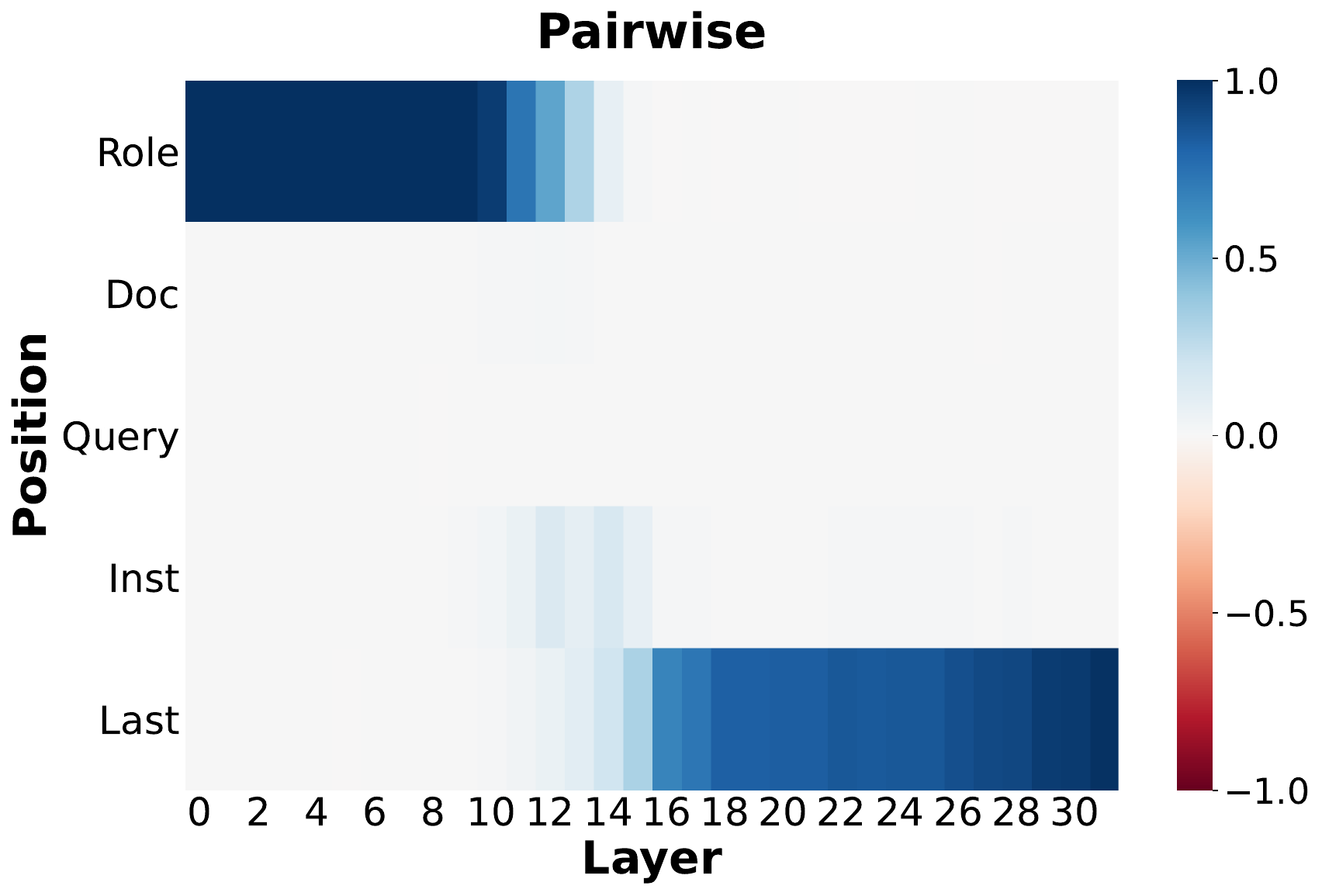}
  \end{subfigure}
  \caption{Results of residual stream patching at layer inputs across prompt segment positions on LLaMa-3.1-8B-Instruct. Darker blue reflects stronger gains.}
  \label{fig:rq1_segment}
\end{figure}

\begin{figure}[t]
  \centering
  \begin{subfigure}{0.49\textwidth}
    \centering
    \includegraphics[width=0.8\textwidth]{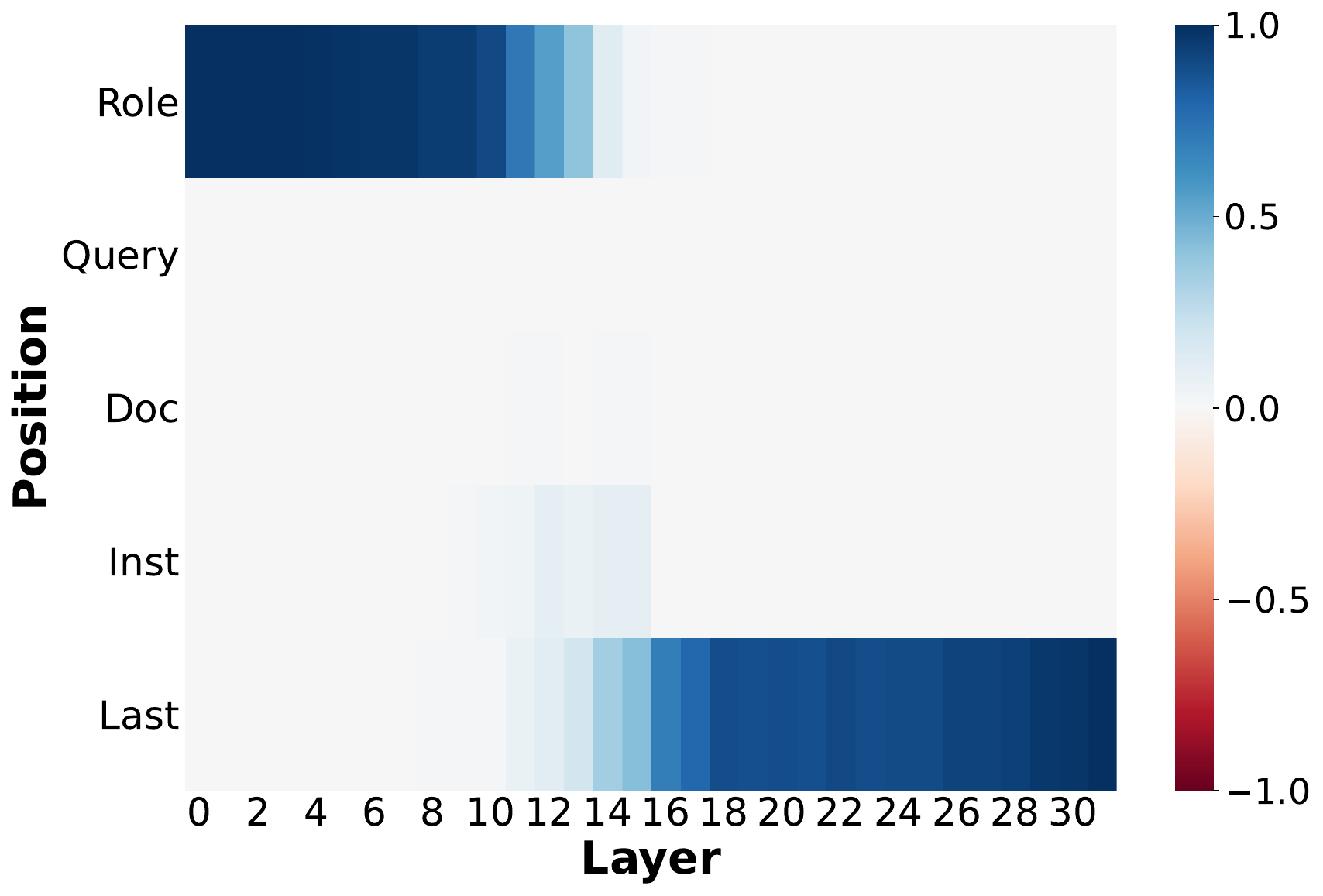}
    \caption{Swapping document and query.}
    \label{fig:rq1_order_1}
  \end{subfigure}
  \hfill
  \begin{subfigure}{0.49\textwidth}
    \centering
    \includegraphics[width=0.8\textwidth]{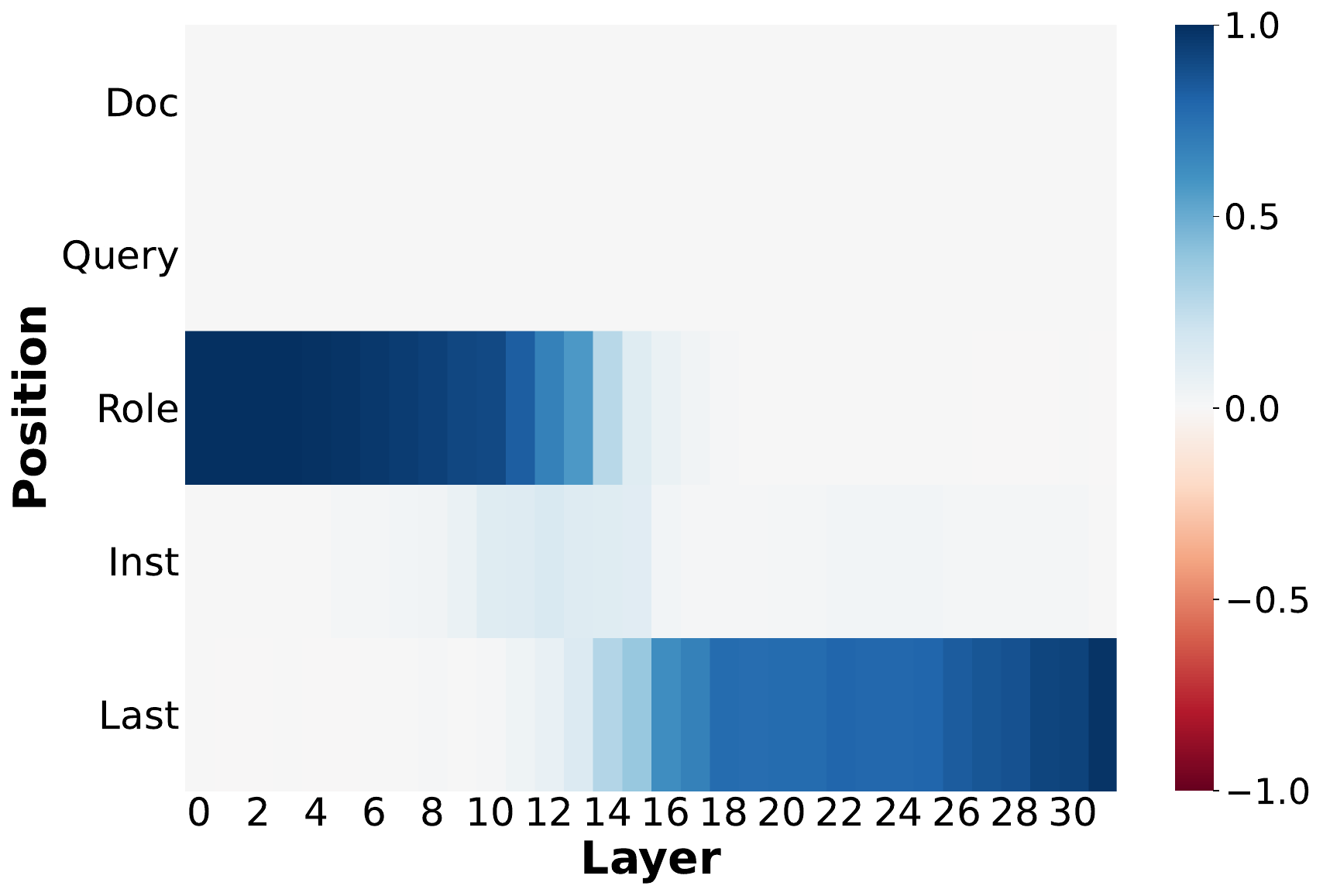}
    \caption{Placing role-play before instruction.}
    \label{fig:rq1_order_2}
  \end{subfigure}
  \caption{Patching result of different
  segment orders 
  for the pointwise ranking.}
  \label{fig:rq1_order}
\end{figure}

Figure~\ref{fig:rq1_segment} shows that \textbf{role-play information originates from role tokens in the lower layers, interacts moderately with instruction tokens in the middle layers, and gradually shifts toward the final token in higher layers}. The mid-layer interaction is expected, as task-related information inevitably leaks into role specifications, making role and instruction difficult to fully disentangle. The shift to the final token is also consistent with prior findings~\cite{liu2025large,meng2022locating}, since ranking scores are derived from the next-token logit at this position~\cite{liu2025large}.
In contrast, patching document or query tokens yields negligible gains in both pointwise and pairwise methods. This implies that role-play signals embedded in the residual stream have limited interaction with query-document representations and are encoded largely independently of task-specific context.

We further examine the effect of prompt ordering on role-play encoding. As shown in Figure~\ref{fig:rq1_order_1}
, swapping the positions of document and query yields negligible changes, and patching these components consistently shows little effect.
In contrast, Figure~\ref{fig:rq1_order_2} shows that the relative position of role and instruction has only modest effects: when role tokens appear at the beginning, their influence remains mostly separable until mid-layers, whereas placing them immediately before the instruction induces slightly earlier role-instruction interaction and weak persistence into higher layers. 
\textbf{In general, the flow of role-play information is largely insensitive to prompt component order}, with only minor mid-layer interactions when the role precedes the instruction.
We observe similar patterns for the Pairwise method.

We additionally examine patching on the outputs of attention and MLP layers. Attention output patching shows similar patterns but with a narrower effect range (-0.2 to 0.2 normalized LD), while MLP patching yields no notable performance changes. We therefore omit detailed results.

\subsection{Attention heads analysis (RQ2)}
\subsubsection{Activation patching on attention heads}\label{sec:attention_head_patching}
\begin{figure}[t]
    \centering
    \includegraphics[width=\linewidth]{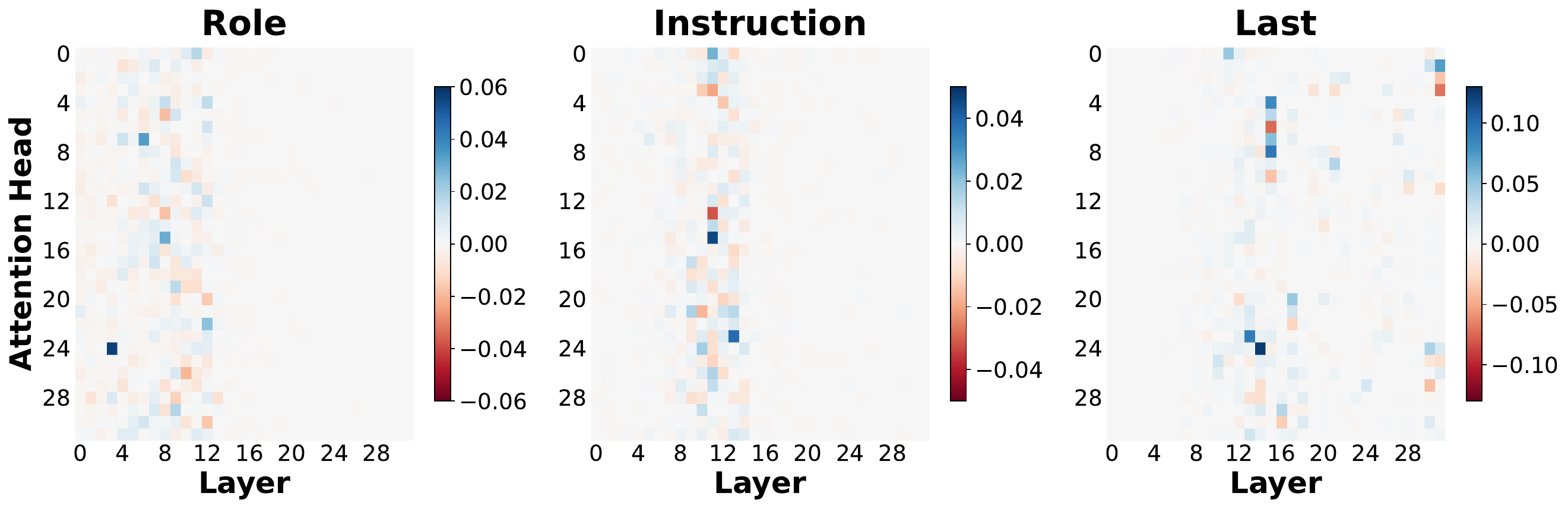}
    \caption{Patching results of attention heads on role, instruction and last token for pointwise ranking.}
    \label{fig:rq2_head}
\end{figure}

We perform attention head patching at different prompt positions, with Figure~\ref{fig:rq2_head} showing representative high-scoring cases.
Overall, the patching scores across positions remain low.
For role tokens, effects concentrate before Layer 13 with L3H24 (0.056) showing a dense effect, while the remaining heads are more sparsely distributed.
Instruction tokens engage several middle-to-lower-layer heads, notably L11H15 (0.046) and L13H23 (0.038).
The last token sees broader and stronger contributions, especially from L14H24 (0.1242) and L13H23 (0.091).
Although these heads causally affect role-conditioned relevance, the relatively small patching scores suggest that their contribution may be weak. To better understand their functional significance in inference, we conduct ablation studies in the next subsection.

\subsubsection{Causal impact of attention heads across tasks}
Specifically, we zero- or mean-ablate~\cite{wang2023interpretability} the top-1 and top-10 attention heads (ranked by patching scores) for each prompt segment position -- role, document, query, instruction, last token -- and for mixed heads across these positions. We report the top-10 results as a representative summary. Ablations are conducted on both pointwise relevance judgment and ranking tasks. 
\begin{figure}[t]
  \centering
  \begin{subfigure}{0.49\textwidth}
    \centering
    \includegraphics[width=\textwidth]{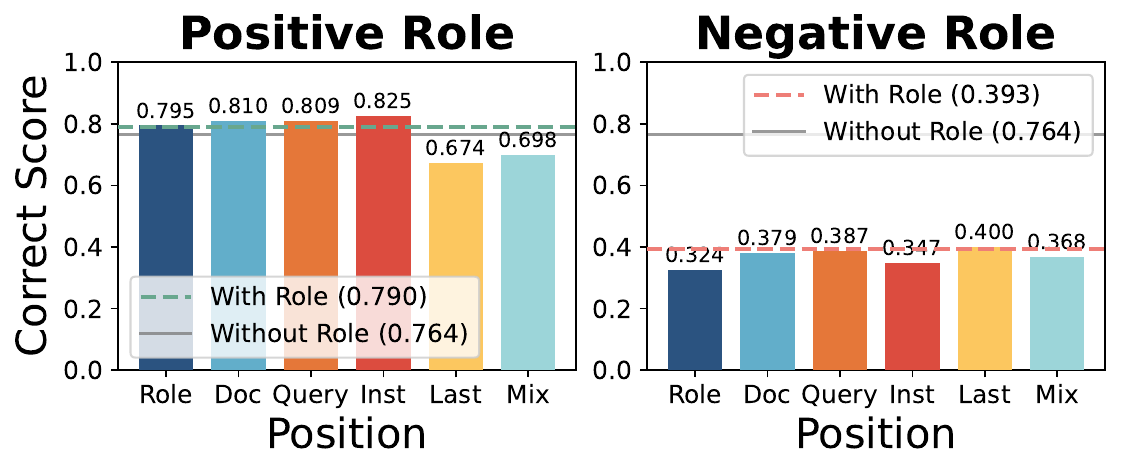}
    \caption{Change in confidence for the correct answer after head ablation, across prompt positions in pointwise ranking.}
    \label{fig:ablation_1}
  \end{subfigure}
  \hfill
  \begin{subfigure}{0.49\textwidth}
    \centering
    \includegraphics[width=\textwidth]{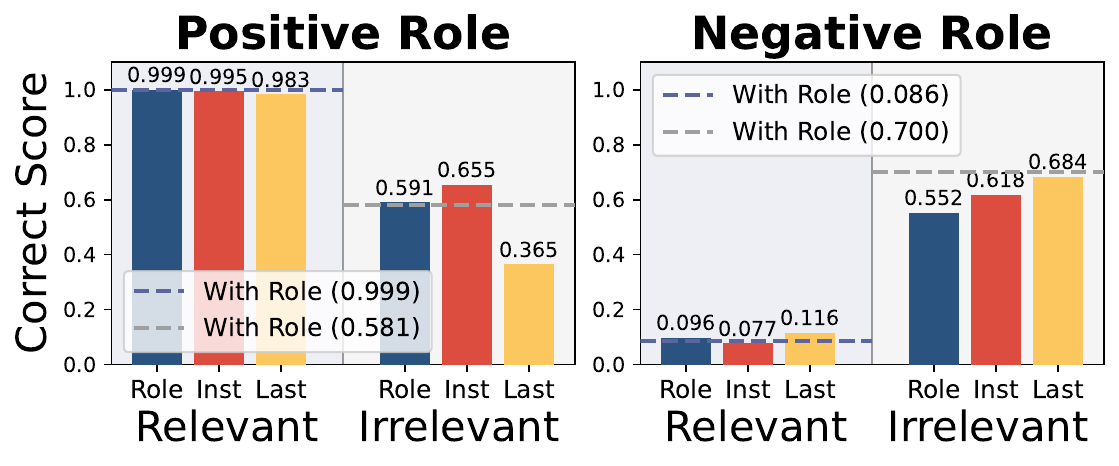}
    \caption{Detailed view of the effect of document types at prompt positions: role, instruction, and last token.}
    \label{fig:ablation_2}
  \end{subfigure}
  \caption{Impact of mean ablation of the top-10 attention heads for each prompt segment identified from the patching results (Section~\ref{sec:attention_head_patching}) on relevance judgment task. Mix indicates ablation of attention heads at other prompt positions.}
  \label{fig:ablation_filtered}
\end{figure}
\paragraph{Relevance judgment evaluation}
This task evaluates the model’s ability to independently assess the relevance of individual query-document pairs.
We use the normalized probability of predicting the correct answer (\textit{correct score}) to reflect model confidence, and this metric is preferred over accuracy, which reduces the task to a binary decision dependent on a threshold.
Correct scores are expected to drop for both role types due to head ablation. 
Figure~\ref{fig:ablation_1} shows that most scores for negative roles do drop ($<0.393$), while some scores for positive roles show a slight increase ($>0.790$). 
To further analyze this unexpected behavior, Figure~\ref{fig:ablation_2} compares performance changes across different document types for positions of interest (role, instruction, and last token).
We find that, \textbf{for both role types, ablation effects are primarily driven by behavior on irrelevant documents}. 
And the failure drop of irrelevant documents likely reflects the inherent difficulty in defining irrelevance consistently, as documents can be irrelevant in multiple ways. 


\begin{figure}[t]
  \centering
  \begin{subfigure}{0.49\textwidth}
    \centering
    \includegraphics[width=\textwidth]{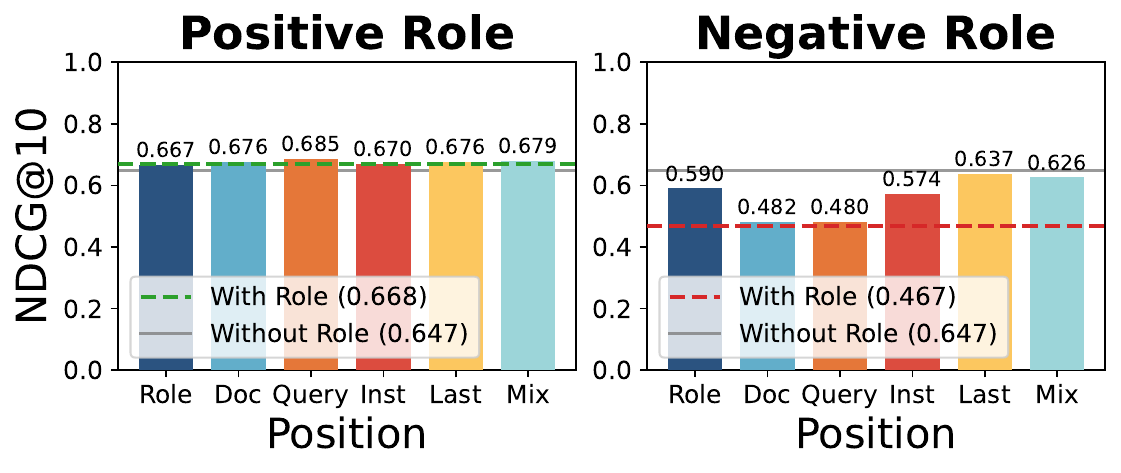}
    \caption{Change in NDCG@10 across prompt positions in pointwise ranking.}
    \label{fig:ablation_3}
  \end{subfigure}
  \hfill
  \begin{subfigure}{0.49\textwidth}
    \centering
    \includegraphics[width=\textwidth]{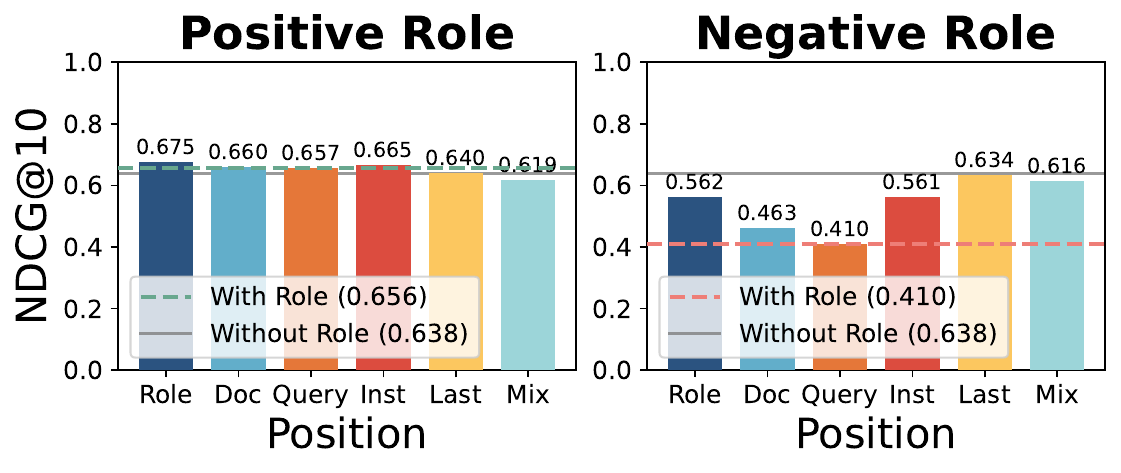}
    \caption{Change in NDCG@10 across prompt positions in pairwise ranking.}
    \label{fig:ablation_4}
  \end{subfigure}
  \caption{Impact of mean ablation of the top-10 attention heads for each prompt segment identified from the patching results (Section~\ref{sec:attention_head_patching}) on the ranking task. Mix indicates ablation of attention heads at other prompt positions.}
  \label{fig:ablation_trecdl19}
\end{figure}

\paragraph{Ranking evaluation}
In contrast, the ranking task involves the relative comparison of individual relevance judgments to determine the final order of multiple documents for a given query.
We further evaluate whether heads identified on MSMARCO generalize to TREC DL 2019. Correct scores are used to rerank the top-10 BM25 candidates, with NDCG@10 expected to decrease for positive roles and increase for negative roles. 
Figure~\ref{fig:ablation_trecdl19} shows that the patterns are consistent with the relevance judgment task but more stable, likely because ranking evaluates multiple documents per query and reduces pointwise noise. Overall, the results indicate that the identified heads generalize well from single-document relevance judgments to the multiple-document ranking task.

Across both tasks, ablating even a single head can have a large impact on performance, especially for negative roles.
Position-wise, query and document positions are minimally affected. In contrast, last token heads generally show the strongest effects, suggesting that the model aggregates and finalizes relevance information at the end of the prompt. Mixed positions, which pool heads across segments, lead to more moderated and robust changes.

\subsection{Role-play prompt phrasing strategy (RQ3)}

\begin{figure}[t]
  \centering
  \begin{subfigure}{0.49\textwidth}
    \centering
    \includegraphics[width=0.8\textwidth]{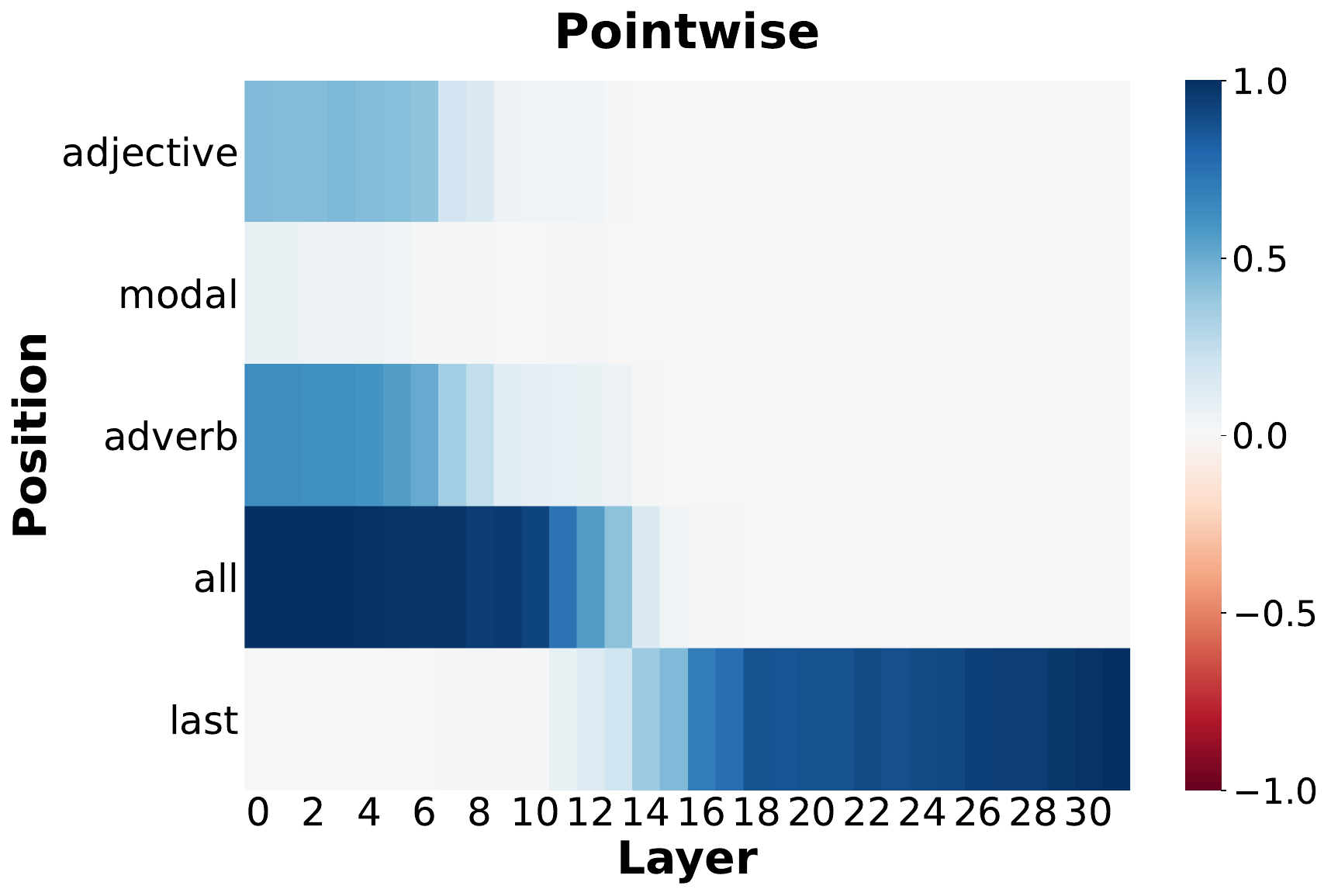}
  \end{subfigure}
  \hfill
  \begin{subfigure}{0.49\textwidth}
    \centering
    \includegraphics[width=0.8\textwidth]{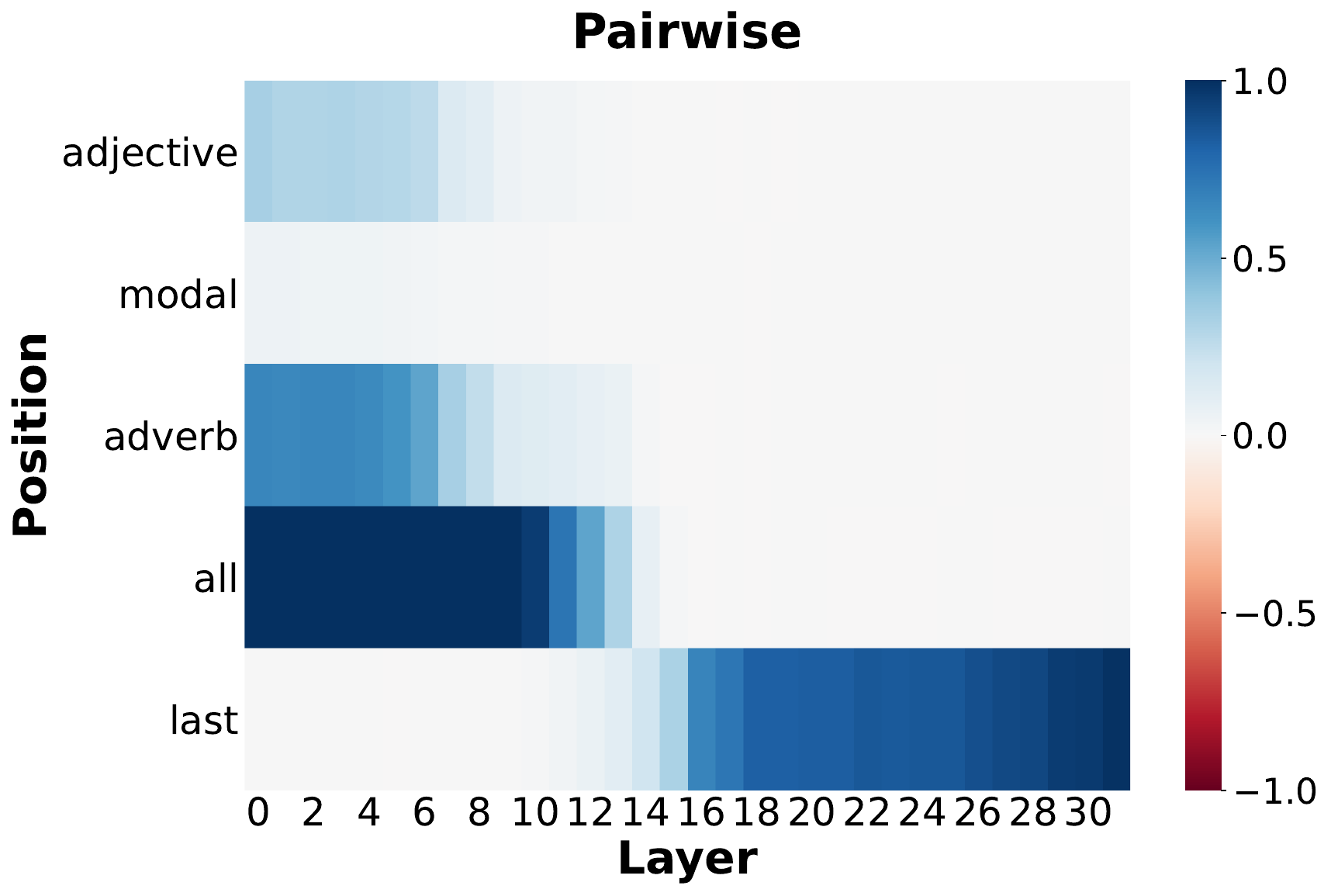}
  \end{subfigure}
  \caption{Results of residual stream patching at layer inputs across different role tokens for both pointwise and pairwise settings. `all' indicates all the tokens in the role-play segment.}
  \label{fig:rq3_role_token}
\end{figure}

Since only three tokens are perturbed in each role-play phrasing, a natural question is which linguistic components contribute most to role-play effectiveness. 
To address this, we present patching results for individual role tokens in Figure~\ref{fig:rq3_role_token}.
Compared to patching all role tokens, a single adverb produces the largest performance improvement, followed by adjective tokens, with modal tokens contributing only minimally. 
These results indicate a clear hierarchy in the importance of different linguistic components: \textbf{In our formulation of the role, the adverb carries the primary role-play signal, the adjective plays a supporting role, and the modal provides only a minor contribution}.
One possible explanation is that the adjective primarily describes the model's capability to perform the task (e.g. ``reliable'' vs ``clumsy''), whereas the adverb actively guides how the model executes the task (e.g. ``accurately'' vs ``wrongly''). In contrast, modal verbs do not encode independent semantic content, but their effect depends on the entire role-play description.
\looseness=-1

These findings provide practical guidance for designing more effective role-play prompts, not only for ranking tasks but also for broader applications in prompt engineering and controlled model behavior.

\subsection{Role-play encoding across models and datasets (RQ4)}
\begin{figure}[t]
    \centering
    \includegraphics[width=\linewidth]{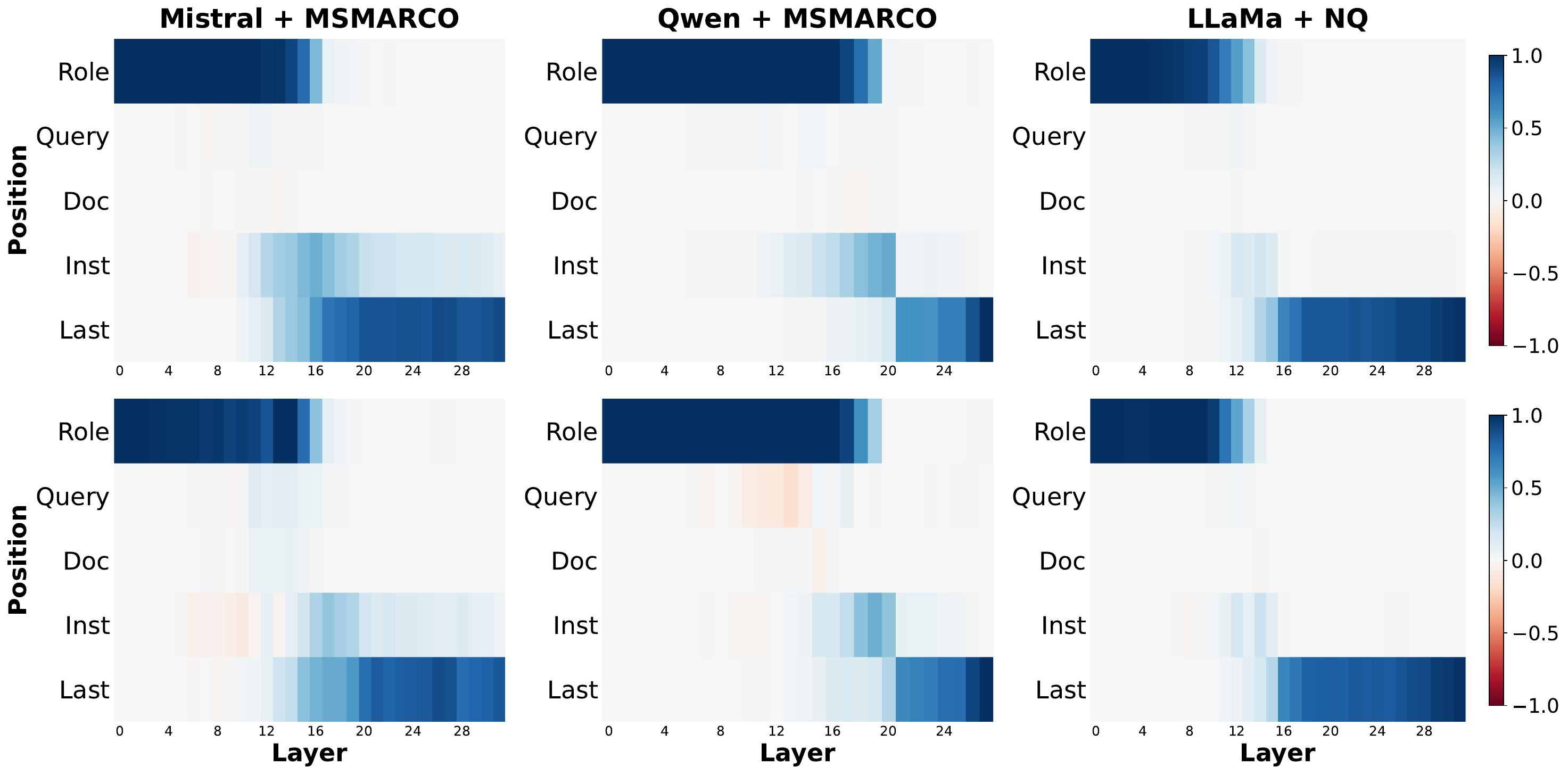}
    \caption{ Results of residual stream patching at layer inputs across different models and datasets. Upper row: pointwise; lower row: pairwise.}
    \label{fig:rq4_model_dataset}
\end{figure}

\subsubsection{Generalization across different models}
We extend activation patching to other instruction-tuned LLMs that share similar scales and architectures: Mistral-7B-Instruct-v0.3~\cite{jiang2023mistral7b} and Qwen2.5-7B-Instruct~\cite{qwen2025qwen25technicalreport}.
As shown in Figure~\ref{fig:rq4_model_dataset}, both models reproduce the core information flow pattern, with role signals gradually propagating to the final token. 
At the same time, model-specific differences emerge: Mistral exhibits earlier and more sustained role-instruction interaction; while Qwen shows a delayed transition to the last token, beginning only around layer 20.
Under the pairwise setting, both models also display slightly stronger but still limited interaction between role and query-document tokens, likely because comparative judgments naturally leverage more contextual signals.
Overall, the results suggest that while role encoding is stable across instruction-tuned LLMs, models differ in the timing and extent of role-instruction interaction, potentially reflecting architectural or training differences.

\subsubsection{Generalization across different datasets}
We also conduct our analyses on the NQ dataset using LLaMa-3.1-8B-Instruct, shown in the last column of Figure~\ref{fig:rq4_model_dataset}. The results reveal patterns highly consistent with those observed on MS MARCO: role signals dominate, interact modestly with instruction tokens, and show minimal entanglement with query-document content.
This consistency suggests that the mechanism underlying role encoding is largely stable across datasets, pointing to an intrinsic property of the model's representation rather than being tightly coupled to any specific data distribution.
\section{Discussion}
While our analysis provides insight into role-play mechanisms, the choice of role descriptors was influenced by activation patching requirements. The need for precise position-wise alignment led us to focus on single-token wordings. Since single-token ranking-related descriptors are scarce, we also included more general or unrelated negative terms (such as `awful' and `sadly'). The fact that such descriptors can still degrade ranking quality indicates that role-play effects need not be semantically tied to ranking expertise, motivating broader mechanistic analysis to prevent unintended or adversarial triggers of ranking degradation.
Our mechanistic analysis provides a roadmap for model steering. Patching reveals that role-play signals are processed early and propagated to the final token in later stages, enabling targeted layer and position interventions, while their weak interaction with query-document representations suggests that role-play can be steered independently without disrupting core relevance, supporting guardrails against negative role leakage and role-based ranking improvements.

\section{Conclusions}

In this paper, we have investigated the effect of role-play in LLM rankers, both through empirical evaluation of positive and negative role-play prompts and through the lens of mechanistic interpretability. 
We found that the formulation and wording of the role-play prompt has a large effect on the zero-shot ranking quality of the LLM. Model activations indicate that role-play signals interact with the instruction part of the prompt and hardly with the query and document representations. Our results generalize across multiple instruction-tuned 8B models and between pointwise and pairwise approaches.
Our findings show that it is important that role-play prompts for LLM rankers are carefully constructed, extending findings from previous work that the effect of prompting can even be larger than the ranking paradigm (pointwise or pairwise).
Our work can be further extended to more models, more datasets, 
and specific domains that may require specific roles.

\begin{credits}
\subsubsection{\ackname}
This publication is part of the project LESSEN with project number NWA.1389.20.183 of the research program NWA ORC 2020/21 which is (partly) financed by the Dutch Research Council (NWO).
\subsubsection{\discintname}
The authors have no competing interests to declare that are relevant to the content of this article.
\end{credits}



\bibliographystyle{splncs04}
\bibliography{references}

\end{document}